\documentclass[doublecol]{epl2} 
\usepackage{chngcntr}
\usepackage{amsmath}
\title{Mean field theory of assortative networks of phase oscillators}
\author{Juan G. Restrepo\inst{1} \and Edward Ott\inst{2}}
\shortauthor{J. G. Restrepo and E. Ott}
\institute{                    
  \inst{1} Department of Applied Mathematics, University of Colorado, Boulder, Colorado 80309, USA\\
 \inst{2} Institute for Research in Electronics and Applied Physics, University of Maryland, College Park, Maryland 20742, USA
}

\abstract{
Employing the Kuramoto model as an illustrative example, we show how the use of the mean field approximation can be applied to large networks of phase oscillators with assortativity. We then use the ansatz of Ott and Antonsen [Chaos {\bf 19}, 037113 (2008)] to reduce the mean field kinetic equations to a system of ordinary differential equations. The resulting formulation is illustrated by application to a network Kuramoto problem with degree assortativity and correlation between the node degrees and the natural oscillation frequencies. Good agreement is found between the solutions of the reduced set of ordinary differential equations obtained from our theory and full simulations of the system.  These results highlight the ability of our method to capture all the phase transitions (bifurcations) and system attractors. One interesting result is that degree assortativity can induce transitions from a steady macroscopic state to a temporally oscillating macroscopic state through both (presumed) Hopf and SNIPER (saddle-node, infinite period) bifurcations. Possible use of these techniques to a broad class of phase oscillator network problems is discussed. 
}

\begin{document}

\maketitle

\section{Introduction}

Recently there has been much interest in the dynamics of large networks of coupled dynamical units. Such systems are of very broad applicability including such examples as power grids \cite{Carreras}, networks of interacting genes \cite{Kaufman}, neuronal networks \cite{Larremore}, and many others. A key question is that of how topological aspects of the network structure affect the global macroscopic dynamics of the system. In this paper we will emphasize the topological aspects of both degree distribution and (especially) assortativity \cite{Newman} (i.e., the tendency of nodes of a certain type to preferentially link to or avoid linking to nodes of similar type), and we will formulate a mean field approach \cite{Ichinomiya, Ji}, incorporating these topological effects. In particular, we will consider the case in which the dynamical units on each network node are oscillators whose states are specified solely by their respective phases (so-called `phase oscillators'). Thus the amplitudes of the nodal oscillations are fixed and are not dynamically varying. Although there are many phase oscillator models (e.g., neuronal models \cite{So}, models for pedestrians walking on and interacting with a moving foot bridge \cite{Eckhardt}, clapping of hands in large audiences \cite{Taylor}, etc.), for illustrative purposes, we will focus on the particular nodal dynamics and interaction form corresponding to the network Kuramoto problem \cite{Ichinomiya, Restrepo},
\begin{eqnarray}\label{kuramoto}
\frac{d\theta_i}{dt} = \omega_i + K\sum_{j=1}^N A_{ij} \sin(\theta_j - \theta_i),
\end{eqnarray}
where the `adjacency matrix' elements $A_{ij}$ are either $0$ or $1$. Equation (\ref{kuramoto}) is a generalization of the original globally coupled ($A_{ij} = 1$ for all $i$ and $j$) Kuramoto model \cite{Kuramoto}.

In this paper we formulate a mean field approximation for Eq.~(\ref{kuramoto}). Our formulation generalizes the mean field formulation of Ref.~\cite{Ichinomiya} to include directed networks, correlation between node degree and frequency and, most importantly, assortativity. The mean field equations that result are ostensibly very difficult to solve. However, we will show  that the ansatz of Ref.~\cite{Ott1} can be employed to reduce the mean field microscopic description for the probability distribution of the model states to an exact macroscopic description of the long-time \cite{Ott2} system dynamics (a finite set of ordinary differential equations) in terms of a set of `order parameters' \cite{footnote}.

As an illustration of our formulation  we consider the network Kuramoto problem with correlation between the network degree and the node frequencies \cite{Skardal1,Skardal2} and with degree assortativity. An important result from this example is that degree assortativity can induce phase transitions from a steady macroscopic state to a temporally oscillating macroscopic state. 

Again we emphasize that the general type of formulation used here can be employed and generalized to treat other situations involving large phase oscillator networks. 

\section{Mean field formulation}\label{sec2}

We consider a random network of $N \gg 1$ nodes. The network is constructed as follows. There is a given degree distribution $P({\bf k})$, ${\bf k} = (k^{in},k^{out})$, which specifies the number of nodes that have $k^{in}$ directed links into them and $k^{out}$ directed links out from them [note that $P$ is normalized such that $\sum_{{\bf k}}P({\bf k}) = N$]. There is a given  frequency probability distribution $g(\omega|{\bf k})$ which in general can depend on the node degree ${\bf k}$. Finally, there is a given assortativity function $a({\bf k}' \to {\bf k})$ which specifies the probability that two nodes of degrees ${\bf k}'$ and ${\bf k}$ are connected by a link from the node of degree ${\bf k}'$ to the node of degree ${\bf k}$. Denoting the average degree $\langle k \rangle = \sum_{{\bf k}} k^{in} P({\bf k}) = \sum_{{\bf k}} k^{out} P({\bf k})$, the total number of network links is $N \langle k \rangle$. Thus the assortativity function is constrained to satisfy
\begin{eqnarray}\label{constraint}
\sum_{{\bf k'}} \sum_{{\bf k}} P({\bf k}')a({\bf k}' \to {\bf k}) P({\bf k}) = N \langle k \rangle.
\end{eqnarray}
In addition, since $a({\bf k}' \to {\bf k})$ is a probability it is also constrained to satisfy $0 \leq a({\bf k}' \to {\bf k}) \leq 1$ when $P({\bf k})P({\bf k}') > 0$. In the absence of assortativity, the probability of a link from node $j$ to node $i$ is proportional to the out-degree from $j$ and the in-degree to $i$, which by (\ref{constraint}) yields
\begin{eqnarray}\label{uncorrelated}
a({\bf k}_j \to {\bf k}_i) = \frac{k^{out}_j k^{in}_i}{N\langle k \rangle}.
\end{eqnarray}
[However, we will be especially interested in cases where (\ref{uncorrelated}) does not hold.]

The random network is formed by first assigning degrees ${\bf k}$ to each node according to the degree distribution $P({\bf k})$. Then each node is randomly assigned a natural oscillation frequency according to the distribution $g(\omega|{\bf k})$. Finally, $a({\bf k}' \to {\bf k})$ is used to form the links between nodes.

In the mean field treatment we approximate the condition $1\ll N < \infty$ by adopting the $N\to \infty$ limit and assuming that the complete network state can be specified by a smoothly varying distribution function $f(\theta, \omega | {\bf k}, t)$ which is defined so that $f d\omega d\theta /(2\pi)$ is the probability at time $t$ that a node of degree ${\bf k}$ has its natural frequency in the range $[\omega, \omega + d\omega]$ and its phase angle in the range $[\theta, \theta + d\theta]$. Thus, since a node's natural frequency does not change with time, 
\begin{eqnarray}
\int_0^{2\pi} f d\theta = g(\omega|{\bf k})
\end{eqnarray}
is time independent.

Writing the interaction term in (\ref{kuramoto}) as 
\begin{eqnarray}
\sum_{i=1}^N \mbox{Im}[e^{-i\theta_i} R_i(t)],\label{localr1}\\
R_i(t) = \sum_{j=1}^N A_{ij}e^{i\theta_j},\label{localr2}
\end{eqnarray}
we identify the nodal order parameters $R_i(t)$ with an assumed mean-field order parameter $R({\bf k},t)$ via $R_i(t) \to R({\bf k}_i,t)$, and we conjecture that this identification provides a good approximation for the macroscopic network dynamics when the nodal degrees are large. From Eq.~(\ref{localr2}) we have
\begin{eqnarray} \label{meanlocalr}
R({\bf k},t) = \sum_{{\bf k}'} P({\bf k}') a({\bf k}' \to {\bf k}) \int \int f(\theta',\omega'|{\bf k}',t) e^{i\theta'}\frac{d\theta}{2\pi}'d\omega'.
\end{eqnarray}
In addition, by the continuity of phase space density, Eqs.~(\ref{kuramoto}) and (\ref{localr2}) yield
\begin{eqnarray}\label{continuity}
&\frac{\partial }{\partial t}f(\theta,\omega|{\bf k},t) +\nonumber\\
& \frac{\partial}{\partial \theta}\left\{ [ \omega + K \mbox{Im}(e^{-i \theta} R({\bf k},t))] f(\theta,\omega|{\bf k},t) \right\} = 0. 
\end{eqnarray}
Equations (\ref{meanlocalr}) and (\ref{continuity}) constitute the mean field approximation to the Kuramoto network model on a directed network with degree assortativity and correlation between the nodal degree ${\bf k}$ and the natural oscillation frequency $\omega$. In the special case of an undirected network, $A_{ij} = A_{ji}$, the mean field formulation is simply obtained by replacing the two component vector degree ${\bf k} = (k^{in},k^{out})$ by the scalar degree $k$ (our numerical example will be for the undirected case). 

\section{Model Reduction}\label{sec3}

The mean-field equations (\ref{meanlocalr}) and (\ref{continuity}) are still difficult to solve in general. Thus, to make further progress, we restrict our attention to the long time asymptotic dynamics of the system. That is, we focus on obtaining the attractors and bifurcations of the mean field system. For this purpose, the results of Refs.~\cite{Ott1} and \cite{Ott2} imply that, in the long time asymptotic limit, the distribution $f$ tends to the special form
\begin{eqnarray}\label{ansatz}
f(\theta,\omega|{\bf k},t) = \left\{  1+ \left[   \sum_{n=1}^{\infty}(b(\omega, {\bf k},t))^n e^{-i n \theta} + (\mbox{c.c.})     \right] \right\} g(\omega|{\bf k}),
\end{eqnarray}
where (c.c.)~denotes the complex conjugate of the summation. Substituting the ansatz (\ref{ansatz}) into (\ref{continuity}), we find that (\ref{ansatz}) indeed satisfies (\ref{continuity}) for $b(\omega, {\bf k}, t)$ satisfying
\begin{eqnarray}\label{bseq}
\frac{\partial b}{\partial t} - i \omega b +\frac{K}{2}(R^* b^2 - R) = 0.
\end{eqnarray}
Substituting (\ref{ansatz}) into (\ref{meanlocalr}) we obtain
\begin{eqnarray}\label{eq10}
R({\bf k},t) = \sum_{{\bf k}'} P({\bf k}')a({\bf k}' \to {\bf k}) \int g(\omega'|{\bf k}')b(\omega',{\bf k}',t) d\omega'.
\end{eqnarray}
Equations (\ref{bseq}) and (\ref{eq10}) represent a substantial simplification of the full mean field description as the $\theta$-dependence has been removed from the description.

One could now imagine attacking the system (\ref{bseq}) and (\ref{eq10}) directly (as was done numerically in the globally coupled case in Ref.~\cite{Lafuerza}) or by employing various convenient forms of $g(\omega| {\bf k})$ where the integral over $\omega$ in Eq.~(\ref{eq10}) can be done (e.g., Refs.~\cite{Ott1}, \cite{Lafuerza}, and \cite{Omelchenko}) by evaluating residue contributions at the complex poles of $g(\omega|{\bf k})$. Here we adopt the latter approach and use the simple example of a Lorentzian distribution of natural frequencies,
\begin{eqnarray}
&g(\omega|{\bf k}) = \frac{1}{\pi} \frac{\Delta ({\bf k})}{[\omega - \omega_0({\bf k})]^2 + \Delta^2({\bf k})} \\
 &= \frac{1}{2\pi i}\left\{ \frac{1}{\omega - [\omega_0({\bf k}) + i \Delta({\bf k})]} - \frac{1}{\omega - [\omega_0({\bf k}) - i \Delta({\bf k})]}\right\}.\nonumber
\end{eqnarray}
Following Ref.~\cite{Ott1}, we note that it can be shown that $b(\omega', {\bf k},t)$ is analytic in the upper half $\omega'$-plane where it goes exponentially to zero as $|\omega'| \to \infty$. Thus evaluating the $\omega'$ integral [Eq.~(\ref{eq10})] by the Cauchy residue theorem \cite{Ott1}, inserting the result in Eq.~(\ref{bseq}), and setting $\omega = \omega_0({\bf k}) + i \Delta({\bf k})$, we obtain
\begin{eqnarray}\label{reduced1}
&\left\{ \frac{\partial}{\partial t} + [-i \omega_0({\bf k}) + \Delta({\bf k})]\right\} \hat b({\bf k},t) + \\
&\frac{K}{2} \sum_{{\bf k}'} P({\bf k}') a({\bf k}' \to {\bf k}) [ \hat b({\bf k}',t)^* \hat b^2({\bf k},t) - \hat b({\bf k}',t)] = 0,\nonumber
\end{eqnarray}
where
\begin{eqnarray}\label{reduced2}
\hat b({\bf k},t) \equiv b(\omega_0({\bf k}) + i\Delta({\bf k}),{\bf k},t).
\end{eqnarray}

 \begin{figure}[t]
  \centering
   \includegraphics[width=1\linewidth]{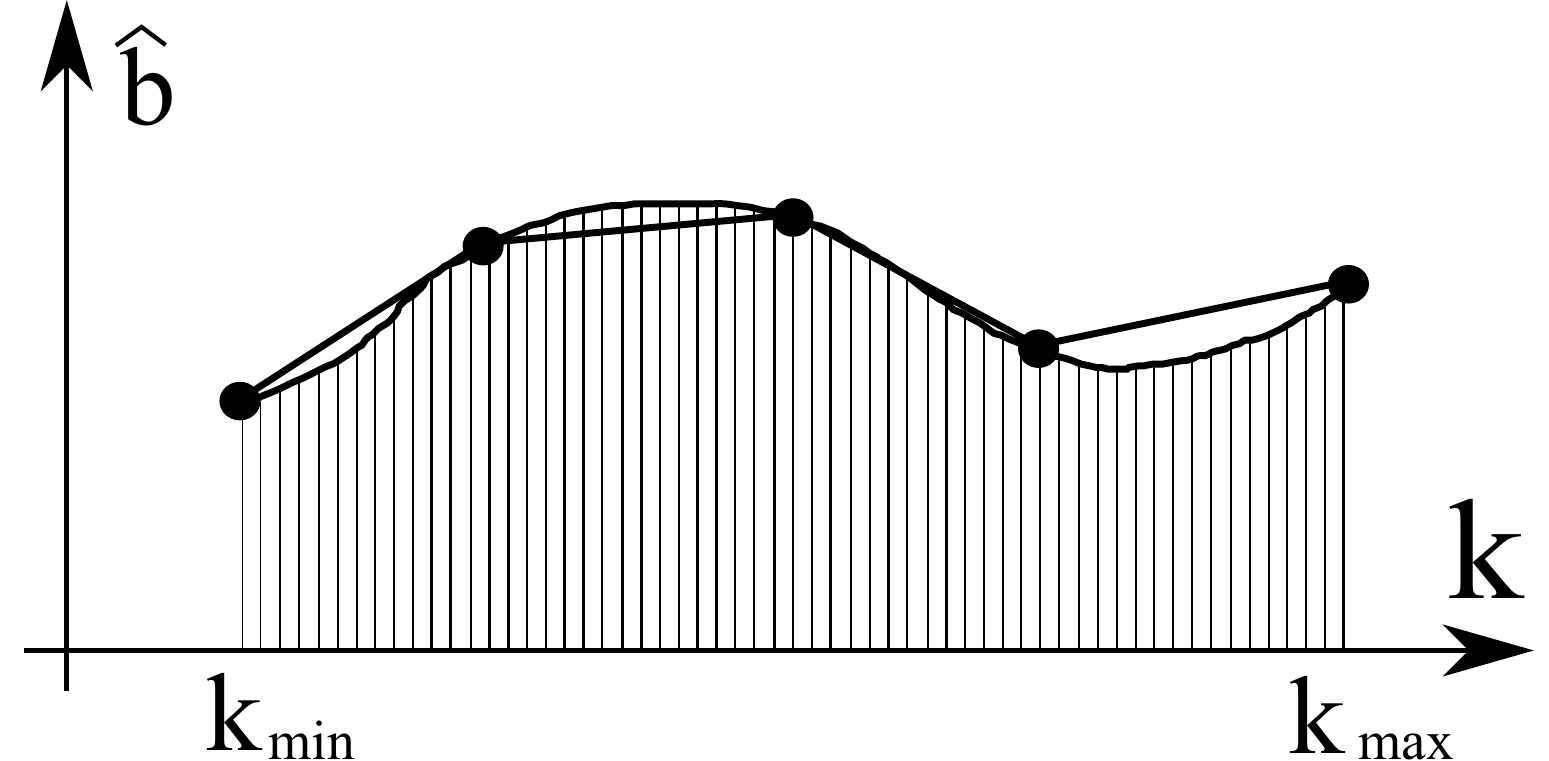}
    \caption{Illustration of a trapezoidal approximation to the order parameter degree spectrum $\hat b(k,t)$ for use in calculating the sums over $k'$ in Eq.~(\ref{reduced1}). Appropriate to an undirected network, in this example $\hat b$ is a function of the scalar degree $k$.}\label{interpolation}
\end{figure}

As compared to Eqs.~(\ref{bseq}) and (\ref{eq10}), Eqs.~(\ref{reduced1}) represent a further substantial reduction. In summary, the original mean field problem [Eqs.~(\ref{meanlocalr}) and (\ref{continuity})] of solving for the macroscopic information [$f(\theta, \omega,{\bf k},t)$] has been {\it exactly} reduced to a closed set of ordinary differential equations for the microscopic variables $\hat b({\bf k},t)$. Compared with the original, finite $N$, Kuramoto problem, Eq.~(\ref{kuramoto}), the system (\ref{reduced1}) has as many equations as there are ${\bf k}$ values, and this can be much further reduced by employing approximation, such as that illustrated in Fig.~\ref{interpolation} for an undirected network case. For the situation illustrated in Fig.~\ref{interpolation}, we envision that $k_{min} \leq k \leq k_{max}$ [$P(k) \equiv 0$ for $k < k_{min}$ or $k > k_{max}$] and that we solve Eq.~(\ref{reduced1}) for the $k$ values $k_{min}$, $k_{max}$, and three intermediate values, with the values of $b(k',t)$ needed for evaluating the sums over $k'$ in (\ref{reduced1}) approximated by interpolating between the five $b(k,t)$ that are explicitly solved for (straight lines in Fig.~\ref{interpolation}).

We next give numerical and analytical examples of the utility of Eq.~(\ref{reduced1}). For our illustration of analytical utility, we take $\omega_0({\bf k}) = \omega_0$ and $\Delta({\bf k}) = \Delta$ [i.e., all nodes have the same $g(\omega)$], and we show how Eq.~(\ref{reduced1}) can be used to simply derive previous results \cite{Restrepo, Restrepo2} for the effects of in/out degree correlation and assortativity on the critical coupling $K = K_c$ at which the incoherent state $\hat b({\bf k},t) = 0$ becomes unstable. Linearizing around $\hat b = 0$ and setting $\hat b({\bf k},t) = \delta({\bf k}) \exp[(i\omega_0 + \gamma)t]$, Eq.~(\ref{reduced1}) yields
\begin{eqnarray}\label{dispersion}
(\gamma + \Delta)\delta({\bf k}) = \frac{K}{2} \mathcal{A}[\delta({\bf k})],
\end{eqnarray} 
where $ \mathcal{A}[\delta({\bf k})]$ denotes the linear operator
\begin{eqnarray}\label{operator}
 \mathcal{A}[\delta({\bf k})] = \sum_{{\bf k}'} P({\bf k}') a({\bf k}' \to {\bf k}) \delta({\bf k}').
 \end{eqnarray}
 Letting $\lambda$ denote the largest real eigenvalue of $\mathcal{A}$, $ \mathcal{A}[\delta] = \lambda \delta$, Eq.~(\ref{dispersion}) yields the critical value $K_c$ at which $\gamma$ goes from negative to positive as $K$ increases through $K_c$,
 \begin{eqnarray}\label{kc}
 K_c = \frac{2\Delta}{\lambda}.
 \end{eqnarray}
   \begin{figure}[t]
  \centering
   \includegraphics[width=0.8\linewidth]{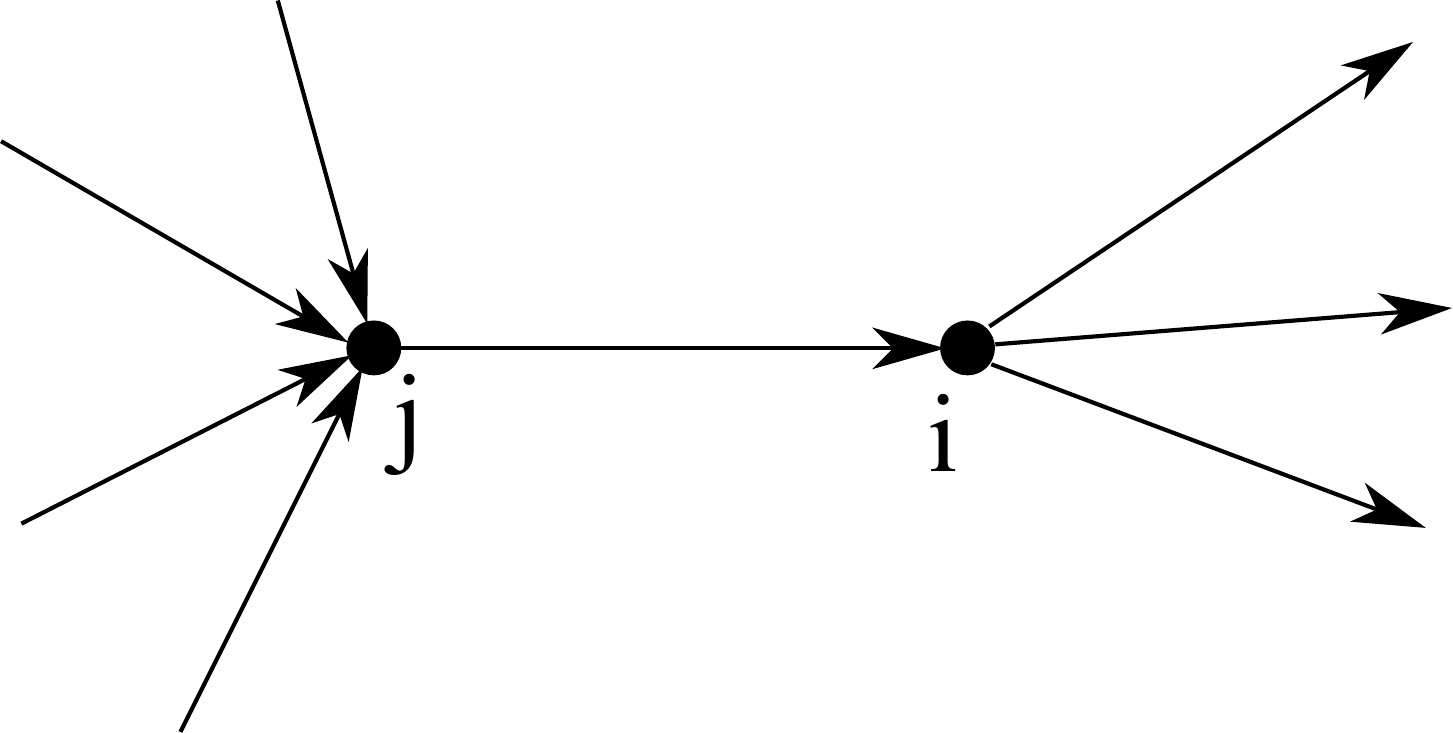}
    \caption{Illustration of the contribution of the edge $j \to i$ to the edge average $\langle k^{in}_j k^{out}_i \rangle_e$, where for the particular edge shown in the figure $k^{in}_j k^{out}_i = 4\times 3 = 12$. }\label{fig:link}
\end{figure}
 Identifying $\lambda$ as the mean field approximation to the largest eigenvalue of the adjacency matrix $[A_{ij}]$, we see that Eq~(\ref{kc}) is in agreement with Eq.~(38) of Ref.~\cite{Restrepo}. Moreover, the eigenvalue problem for $\mathcal{A}$ can be solved by perturbation theory (see Section 1 of the Supplementary Material) to yield
 \begin{eqnarray}\label{lambdarho}
 \lambda \approx \frac{\langle k^{out} k^{in}\rangle}{\langle k \rangle }\rho,
 \end{eqnarray}
 where the assortativity coefficient $\rho$ is defined \cite{Restrepo2} by
 \begin{eqnarray}\label{rho}
 \rho = \frac{\langle k_j^{in} k_i^{out}\rangle_e \langle k \rangle^2  }{\langle k^{in} k^{out}\rangle^2}.
 \end{eqnarray}
 Here $\langle \dots \rangle_e$ indicates an average over all edges $j \to i$ (see Fig.~\ref{fig:link}) which in our mean field description is given by
 \begin{eqnarray}\label{edgeaverage}
 \langle k_j^{in} k_i^{out}\rangle_e = \sum_{{\bf k}} \sum_{{\bf k}'} P({\bf k}') a({\bf k}'\to{\bf k}) P({\bf k}).
 \end{eqnarray}
The assortativity coefficient is one ($\rho = 1$) when there is no assortativity [as may be verified from (\ref{uncorrelated}), (\ref{rho}), and (\ref{edgeaverage})] and is greater (less) than one when the network is assortative (disassortative). Thus (\ref{kc}) and (\ref{lambdarho}) show how $K_c$ is influenced by correlation between the nodal in- and out-degrees (the term $\langle k^{in} k^{out} \rangle$) as well as by assortativity (the factor $\rho$). We note that (\ref{kc}) and (\ref{lambdarho}) have been previously obtained by other methods in Ref.~\cite{Restrepo} and Ref.~\cite{Restrepo2}, respectively.

\section{Numerical Example}\label{sec4}

In this Section we provide a numerical example that illustrates the utility of our approach. We consider an undirected network with $N$ nodes and a degree distribution
\begin{eqnarray}
P(k) = \left\{
\begin{array}{cc}
0, & k < k_{min},\\
C k^{-\gamma}, & k_{min}\leq k < k_{max},\\
0, & k_{max} \leq k,\label{pk}
\end{array}
\right.
\end{eqnarray}
where $C$ is chosen so that $\sum_{k = k_{min}}^{k_{max}}P(k) = N$. An undirected link is established between a node $j$ with degree $k_j$ and a node $i$ with degree $k_i$ with probability
\begin{eqnarray}
&a(k_j \to k_i) = h(a_{ij})
\end{eqnarray}
where
\begin{eqnarray}
&a_{ij} = \frac{k_i k_j}{N \langle k \rangle}\left[ 1 + c \left(\frac{k_i - \langle k \rangle}{ k_i} \right) \left(\frac{k_j - \langle k \rangle}{k_j} \right)  \right],
\end{eqnarray}
$h(x) = \max(\min(x,1),0)$ guarantees $0\leq a(k_j \to k_i)  \leq 1$, and $c$ is a parameter used to adjust the degree of assortativity. In the majority of our simulations $a_{ij}$ satisfies $0 \leq a_{ij} \leq 1$. In this case, $a(k_j \to k_i) = a_{ij}$ satisfies the constraint (\ref{constraint}), and the expected degree of a node $i$ over network realizations is $k_i$, as can be checked by taking the expected value of $\sum_{j =1}^N A_{ij}$, where $A_{ij} = 1$ with probability $a_{ij}$ and $0$ otherwise. Moreover, in this case $c$ and the assortativity coefficient $\rho$ in Eq.~(\ref{rho}) are related by 
\begin{eqnarray}
\rho = 1 + c\left(\frac{\langle k^2 \rangle - \langle k \rangle^2}{\langle k^2 \rangle}\right)^2.
\end{eqnarray}

In networks with high assortativity (high $c$), there can be a small fraction of pairs $i$, $j$ for which $a_{ij} < 0$. In this case, after the links are established the resulting degree distribution may be slightly different than the original target degree distribution $P(k)$, and therefore we will use the realized distribution $\tilde P(k)$ in our reduced theory [Eq.~(\ref{reduced1})] instead of the original target distribution (\ref{pk}). Note that in this case $\tilde P({\bf k})$ and $a(k_j \to k_i)$ satisfy (\ref{constraint}).

For our example, we choose $N = 5000$, $k_{min} = 50$, $k_{max} = 300$, and $\gamma = 3.0$. We take the distribution of frequencies for nodes of degree $k$ to be a Lorentzian with mean $\omega_0(k) = 0.05k$ and width $\Delta =  1$.  We construct undirected networks for different values of the assortativity $\rho$ (corresponding to different values of $c$) as described above.   

We find that for $\rho = 1$ or $\rho < 1$ (disassortative) as $K$ increases from zero there is a bifurcation from incoherence to a macroscopic steady state which, as in the original Kuramoto model, persists as $K$ is further increased. In contrast, for sufficiently large assortativity, we find the surprising result that bifurcations between steady and oscillatory states become possible. In order to illustrate this, in what follows, we focus on the case $\rho = 1.15$.  (See also Sec. 2 of the Supplementary Material)

We integrate (\ref{kuramoto}) numerically using an Euler scheme with $\Delta t = 0.002$ with the phases initially distributed uniformly in $[0,2\pi)$ and $\hat K \equiv 50K = 1.0$, and we increase $\hat K$ by $0.1$ every $50$ time units.
We calculate the time average of the order parameter
\begin{eqnarray}\label{rt}
R(t) = \frac{\left| \sum_{i=1}^N R_i(t)\right| }{N\langle k \rangle} = \frac{\left| \sum_{i,j} A_{ij} e^{i\theta_j}\right| }{\sum_{i,j} A_{ij} }
\end{eqnarray}
where $R_i(t)$ is defined in Eq.~(\ref{localr2}), and store a time series of $R(t)$. Note that the order parameter $R(t)$ is approximately $0$ if the phases are uniformly distributed in $[0,2\pi)$ and $1$ if they are equal.

In addition to numerically solving Eqs.~(\ref{kuramoto}), we numerically solve the reduced system (\ref{reduced1}) using an analogous protocol, i.e., we choose small but nonzero initial conditions $\hat b(k,0) = 0.01$, set the coupling constant $\hat K$ initially to $1.0$, and increase it by $0.1$ every $t = 50$ units. From the solution of Eqs.~(\ref{reduced1}), we compute the order parameter
\begin{eqnarray}\label{bt}
&B(t) = \frac{1}{N\langle k \rangle}|\sum_{k}P(k) R(k,t)| \nonumber \\
&=\frac{1}{N\langle k \rangle}|\sum_{k,k'}\tilde P(k) \tilde P(k') a(k' \to k) \hat b(k',t)|.
\end{eqnarray}
Note that when $a(k_i \to k_j) = a_{ij}$ the order parameter simplifies to $B(t) = |\sum_{k} \tilde P(k) k \hat b(k,t)|/(N\langle k \rangle)$.

\begin{figure}[t]
\setlength{\belowcaptionskip}{-10pt}
  \centering
   \includegraphics[width=1\linewidth]{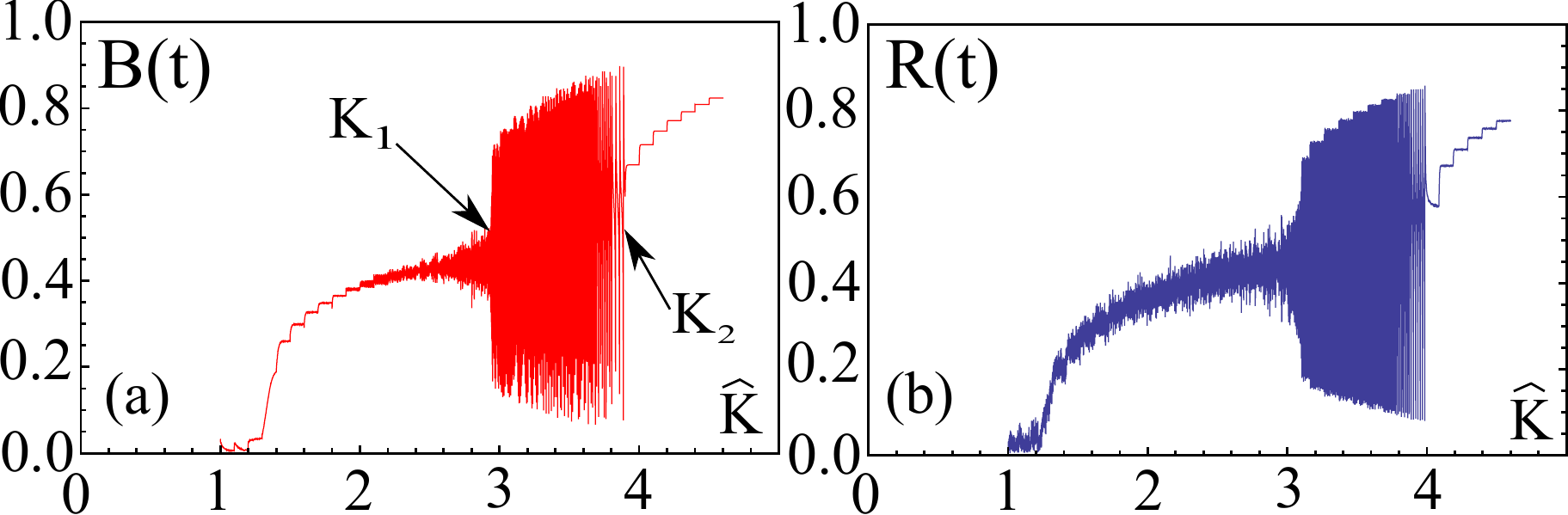}
    \caption{(a) Order parameter $B(t)$ in Eq.~(\ref{bt}) calculated directly from Eqs.~(\ref{reduced1}). (b) Order parameter $R(t)$ in Eq.~(\ref{rt}) calculated using the reduced equations (\ref{kuramoto}). Note that in the text we defined $\hat K  \equiv 50 K $.}\label{compare1}
\end{figure}

In Figs.~\ref{compare1}~(a) and (b) we plot $B(t)$ and $R(t)$, respectively,  obtained from these simulations. We note that there is very good agreement between the simulations  of the full model and of the reduced system. A remarkable behavior observed both in the simulations of the full model and in the reduced system is a transition with increasing $K$ from a steady synchronized state to a temporally oscillating macroscopic state and a subsequent transition to another steady synchronized state. In Fig.~\ref{compare1} (a) we see that around $\hat K \approx 1.3$ there is a transition from incoherence ($B(t) \approx 0$) to a steady synchronized state ($B(t) = B > 0$). As $K$ is increased, the order parameter $B(t)$ develops small oscillations, and these oscillations increase in amplitude as $K$ increases. At some value $K = K_1$, the amplitude of oscillations increases discontinuously. As $K$ is increased further, the period of the oscillations increases until at another value $K = K_2$ the period diverges and the oscillations disappear.

\begin{figure}[t]
\setlength{\belowcaptionskip}{-10pt}
  \centering
   \includegraphics[width=1\linewidth]{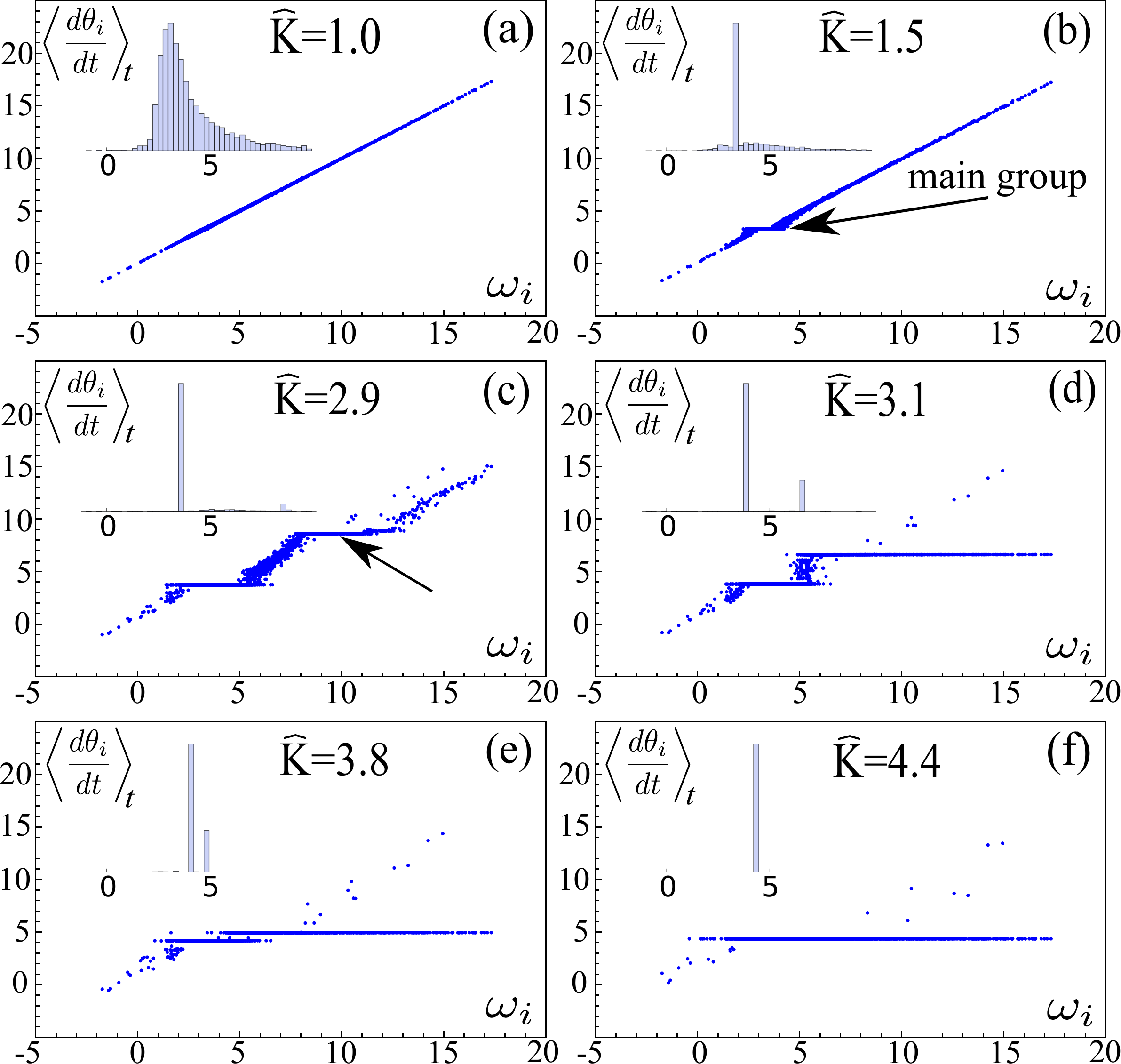}
    \caption{ Time-averaged effective frequency $\langle d\theta_i/ dt \rangle_t$ versus the intrinsic frequency $\omega_i$ for $\hat K = 1.0, 1.5, 2.9, 3.1, 3.8$, and $4.4$. The inset shows a histogram of the time-averaged effective frequencies.}\label{panel2}
\end{figure}

To gain insight into the sequence of bifurcations, we plot in Fig.~\ref{panel2} the time-averaged effective frequency $\langle d\theta_i/ dt \rangle_t$ versus the intrinsic frequency $\omega_i$ for $\hat K = 1.0, 1.5, 2.9, 3.1, 3.8$, and $4.4$.
The insets show histograms of the effective frequencies $\langle d\theta_i/ dt \rangle_t$. At $\hat K = 1.0$ [Fig.~\ref{panel2} (a)], the oscillators are incoherent and the effective frequencies correspond to the intrinsic frequencies, which are distributed according to $\int P(k) g(\omega | k) dk$. At $\hat K = 1.5$ [Fig.~\ref{panel2} (b)], a group of oscillators (labeled `{\it main group}', and indicated by an arrow) has locked at a common frequency, resulting in a macroscopic synchronized steady state. At $\hat K = 2.9$ [Fig.~\ref{panel2} (c)], another group of oscillators (indicated by an arrow) with higher degrees have locked at a higher frequency, which results in oscillations of the order parameter at the difference between the frequencies of the two locked groups. Since the number of oscillators in the second locked group is relatively small [see inset to Fig.~\ref{panel2} (c)], the resulting oscillations in the order parameter are small.  As expected, we observe that the frequency of oscillations  of $R(t)$ is the difference between the values of $\langle d\theta_i/ dt \rangle_t$ for the two locked groups. At $\hat K = 3.1$ [Fig.~\ref{panel2} (d)], most of the oscillators in the high frequency tail of the distribution have locked to a common frequency, while the oscillators with low frequencies remain locked to another frequency. This results in an increased amplitude of the oscillations. At $\hat K = 3.8$ [Fig.~\ref{panel2} (e)], the frequencies have become closer, which results in a smaller frequency in the oscillations of the order parameter. As $\hat K \to K_2$, the frequencies approach each other and the oscillation period of the  order parameter diverges, until almost all the oscillators lock to a common frequency as shown for $\hat K = 4.4$ in Fig.~\ref{panel2} (f). We note that a similar bifurcation scenario, with multiple synchronized clusters inducing oscillations of the order parameter, was recently observed in the Kuramoto model with inertia in Ref.~\cite{Olmi}.

\begin{figure}[t]
\setlength{\belowcaptionskip}{-10pt}
  \centering
   \includegraphics[width=1\linewidth]{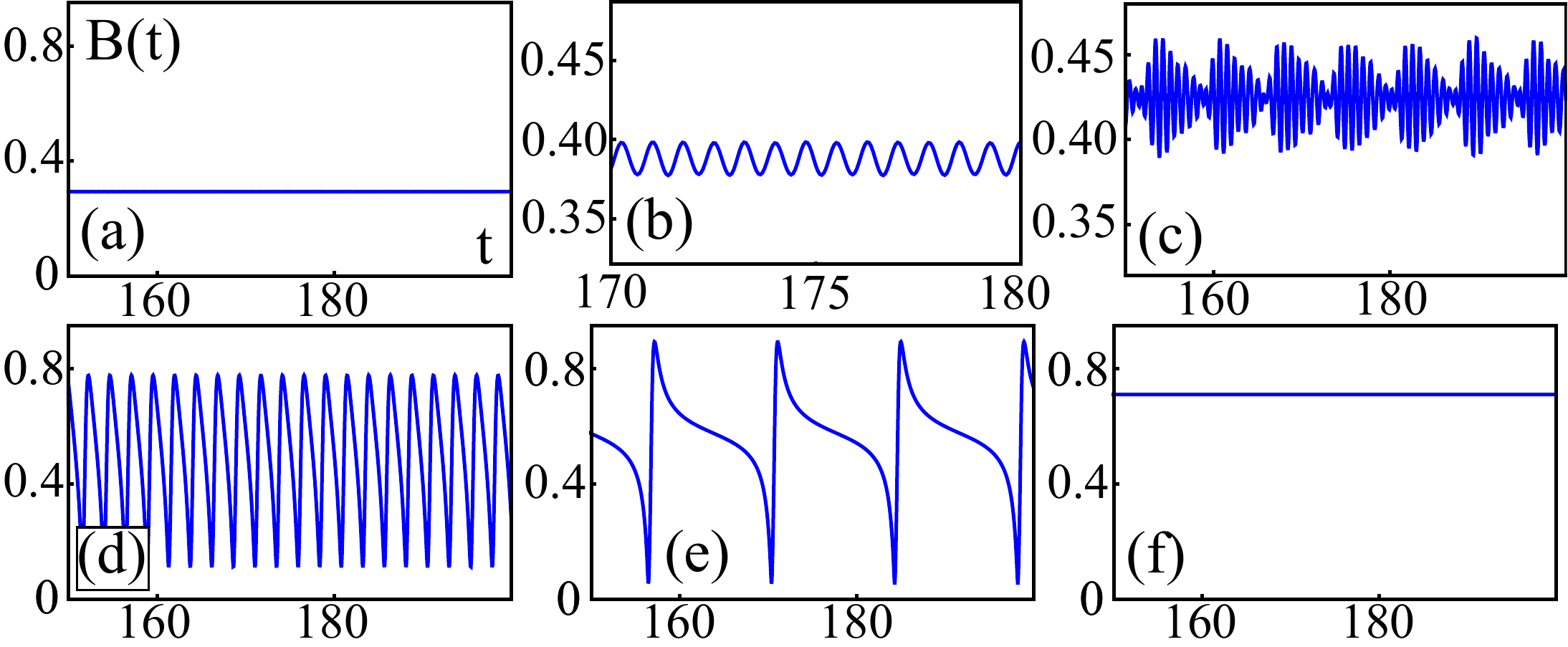}
    \caption{Order parameter $B(t)$ as a function of time for $\hat K = 1.5$ (a), $2.0$ (b), $2.5$ (c), $3.1$ (d), $3.8$ (e), and $4.4$ (f). Note the different scales used in Figs.~(b) and (c).}\label{series}
\end{figure}

We interpret the sequence of bifurcations as follows. The steady synchronized state appears when a group of oscillators (which we will call {\it the main group}) synchronizes with a common frequency. The bifurcation leading to the order parameter oscillations around this steady synchronized state appears through a (presumed) Hopf bifurcation, when a new group of oscillators locks to a frequency separate from the frequency of the main group. Since the number of oscillators in this group can be small, the amplitude of these order parameter oscillations can be small. We also observe that these periodic oscillations become quasiperiodic [Fig.~\ref{series} (c)] as additional groups of oscillators lock to separate frequencies. At $K = K_1$, the oscillations become large and periodic as a large locked group (separate from the main group) is abruptly formed. As $K$ is increased the frequency difference between the two synchronized groups decreases, leading to a (presumed) SNIPER (saddle-node, infinite-period) bifurcation at $K = K_2$. In Fig.~\ref{series} we illustrate the steady (a), low-amplitude periodic (b), quasiperiodic (c), large-amplitude periodic with small period (d), large-amplitude periodic with large period (e), and steady (f) behaviors by plotting $B(t)$ as a function of $t$ for $\hat K = 1.5, 2.0, 2.5, 3.1, 3.8,$ and $4.4$. We note here that the reduced equations (\ref{reduced1}) allow us to visualize the small and quasiperiodic oscillations in the order parameter [Figs.~\ref{series} (b) and (c)] which are masked in the time series of $R(t)$ by noise due to finite size effects.

For other choices of degree distributions and parameters, we have observed qualitatively different bifurcations such as discontinuous transitions from one steady synchronized state to another, or bifurcations like those described above, but in which the mean of the oscillations changes discontinuously as $K$ is increased. In the Supplementary Material we include an additional example which illustrates how the novel sequence of bifurcations appears for assortative networks. Since our goal here is only to illustrate the applicability and usefulness of our mean field approach, we postpone a more detailed study of additional cases for future research. As further indication of the usefulness of the reduced system, we note that the bifurcations presented above were observed first by solving the reduced equations, and were later confirmed by solving the full system (\ref{kuramoto}).

We note that in some cases we have observed discrepancies between the simulations of the full and the reduced system in the values of $K$ at which the onset of oscillations occurs or in the amplitude of these oscillations. One example of this is shown in the Supplementary Material. We speculate that these differences might be due to a combination of (i) insufficient number of nodes of a given degree $k$ (especially of large degree) for the mean-field assumption to be valid and to provide a good sample of $g(\omega| {\bf k})$, (ii) finite size fluctuations driving the system away from the manifold where (\ref{ansatz}) holds, and/or (iii) sensitivity of Eqs.~(\ref{reduced1}) to $\tilde P(k)$ [we have noted, for example, that results can be slightly different when using the target distribution $P(k)$ instead of the realized distribution $\tilde P(k)$].

\section{Conclusion}\label{sec5}

Our main general conclusion is that the combined use of mean field theory and the ansatz of Ref.~\cite{Ott1} provides a very promising technique for exploring and discovering topological network effects on the dynamics of large interconnected phase oscillator systems with both directed and undirected links. Such topological effects include degree distribution, nodal correlations between in and out degrees, correlations between nodal frequencies and degrees, and degree assortativity in the formation of links. 

With respect to our illustrative numerical example on a Kuramoto network system, we have shown that topology can have profound and surprising qualitative effects on dynamics. In particular, it was found that assortativity by degree can lead to dynamical transitions of different types between steady state, periodic, and quasiperiodic attractors.

Finally, while our example was for a Kuramoto network system, we emphasize that our technique should be useful in many other contexts. One promising application is to neural networks where Refs.~\cite{So} have developed effective phase oscillator models of neurons and have analyzed systems of such model neurons by use of the ansatz of Ref.~\cite{Ott1}.


\begin{acknowledgments}
The work of E. Ott was supported by ARO Grant W911NF-12-1-0101.
\end{acknowledgments}

\newpage

\begin{centering}
{\Large {\bf Supplementary Material}}
\end{centering}

\section{Eigenvalue of the operator $\mathcal{A}$}\label{apa}

\renewcommand{\thefigure}{S\arabic{figure}}

\setcounter{figure}{0}

In order to solve $\mathcal{A}[\delta({\bf k})] = \lambda \delta({\bf k})$ for the case with assortativity, we expand $a({\bf k}' \to {\bf k})$ for small assortativity
\begin{equation}\tag{S1}
a({\bf k}' \to {\bf k}) = a^{(0)}({\bf k}' \to {\bf k}) + a^{(1)}({\bf k}' \to {\bf k}),\label{a1} \\
\end{equation}
\begin{equation}
a^{(0)} \gg |a^{(1)}|,\nonumber
\end{equation}
with $a^{(0)}$ given by the non assortative link probability specified by Eq.~(3). Correspondingly, we expand $\lambda$ and $\delta({\bf k})$ as 
\begin{equation}
\lambda = \lambda^{(0)} + \lambda^{(1)},\label{a2}\tag{S2}
\end{equation}
\begin{equation}
\delta({\bf k}) = \delta^{(0)}({\bf k}) + \delta^{(1)}({\bf k})\label{a30}\tag{S3},
\end{equation}
with $|\lambda^{(0)}| \gg |\lambda^{(1)}|$ and $|\delta^{(0)}({\bf k})| \gg |\delta^{(1)}({\bf k})|$. To lowest order (S1)-(S3) and (3) yield 
\begin{equation}\tag{S4}
\lambda^{(0)} \delta^{(0)}({\bf k}) = \frac{k^{in}}{N\langle k \rangle} \sum_{{\bf k}'} (k^{out})'P({\bf k}') \delta^{(0)}({\bf k}').
\end{equation}
Thus $\delta^{(0)}$ is proportional to $k^{in}$, and we obtain
\begin{equation}\tag{S5}
\lambda^{(0)} =\frac{\langle k^{in}k^{out} \rangle}{\langle k \rangle}.
\end{equation}
Proceeding to next order, we have
\begin{equation}\label{a3}
\lambda^{(1)} \delta^{(0)}({\bf k}) + \lambda^{(0)} \delta^{(1)}({\bf k}) =  \frac{k^{in}}{N \langle k \rangle} \sum_{{\bf k}'} (k^{out})' P({\bf k}') \delta^{(1)}({\bf k}') \nonumber\\
\end{equation}
\begin{equation}
+ \sum_{{\bf k}'} P({\bf k}') a^{(1)}({\bf k}' \to {\bf k}) \delta^{(0)}({\bf k}').\tag{S6}
\end{equation}
We now eliminate $\delta^{(1)}({\bf k})$ from (S6) by multiplying through by $k^{out} P({\bf k})$ and summing over ${\bf k}$. This yields
\begin{equation}
\lambda^{(1)} \sum_{{\bf k}}k^{in} P({\bf k}) k^{out} = \label{a4}
\tag{S7}\\
\end{equation}
\begin{equation}
\sum_{{\bf k}} \sum_{{\bf k}'} (k^{in})' P({\bf k}') a^{(1)}({\bf k}' \to {\bf k}) P({\bf k}) k^{out},\nonumber
\end{equation}
which, using $a^{(1)} = a - a^{(0)}$, gives
\begin{equation}\label{a5}
\lambda^{(1)} \langle k^{in} k^{out} \rangle N = \nonumber\\
\end{equation}
\begin{equation}
N \langle k \rangle \langle k_j^{in} k_i^{out}\rangle_e
- \frac{1}{N \langle k \rangle}\left\{  \sum_{{\bf k}} k^{in} P({\bf k}) k^{out} \right\}^2\nonumber 
\end{equation}
\begin{equation}
= N \langle k_j^{in} k_i^{out}\rangle_e - \lambda^{(0)} \langle k^{in} k^{out} \rangle N \tag{S8}.
\end{equation}
Thus Eq.~(S8), with $\lambda = \lambda^{(0)} + \lambda^{(1)}$ and Eq.~(19) for $\rho$, yields Eq.~(18), as claimed.

\section{Additional Numerical Example}

In this Section we provide an additional example showing how our theory is able to capture the different types of bifurcations that occur as the assortativity coefficient is changed. In this example we use $N = 5000$, $k_{min} = 100$, $k_{max} = 300$, and $\gamma = 3.0$.  We choose values of $c$ that correspond to $\rho = 0.96, 0.98,\dots,1.04$. For each value of $\rho$, we construct an undirected network as indicated in the main text.  We take the distribution of frequencies for nodes of degree $k$ to be a Lorenzian with mean $\omega_0(k) = 0.05k$ and width $\Delta =  1$. For each value of $\rho$, we simulate directly Eq.~(1) with the phases initially distributed uniformly in $[0,2\pi)$ and $\tilde K \equiv 100K = 2.5$. Every $t = 200$ time units, we increase $\tilde K$ by $0.05$. In addition to solving Eq.~(1) directly, we numerically solve the reduced system (13) using an analogous protocol, i.e., we choose small but nonzero initial conditions $\hat b(k,0) = 0.01$, set the coupling constant $\tilde K$ initially to $2.5$, and increase it by $0.1$ every $t = 200$ units.

\begin{figure}[h]
  \centering
   \includegraphics[width=\linewidth]{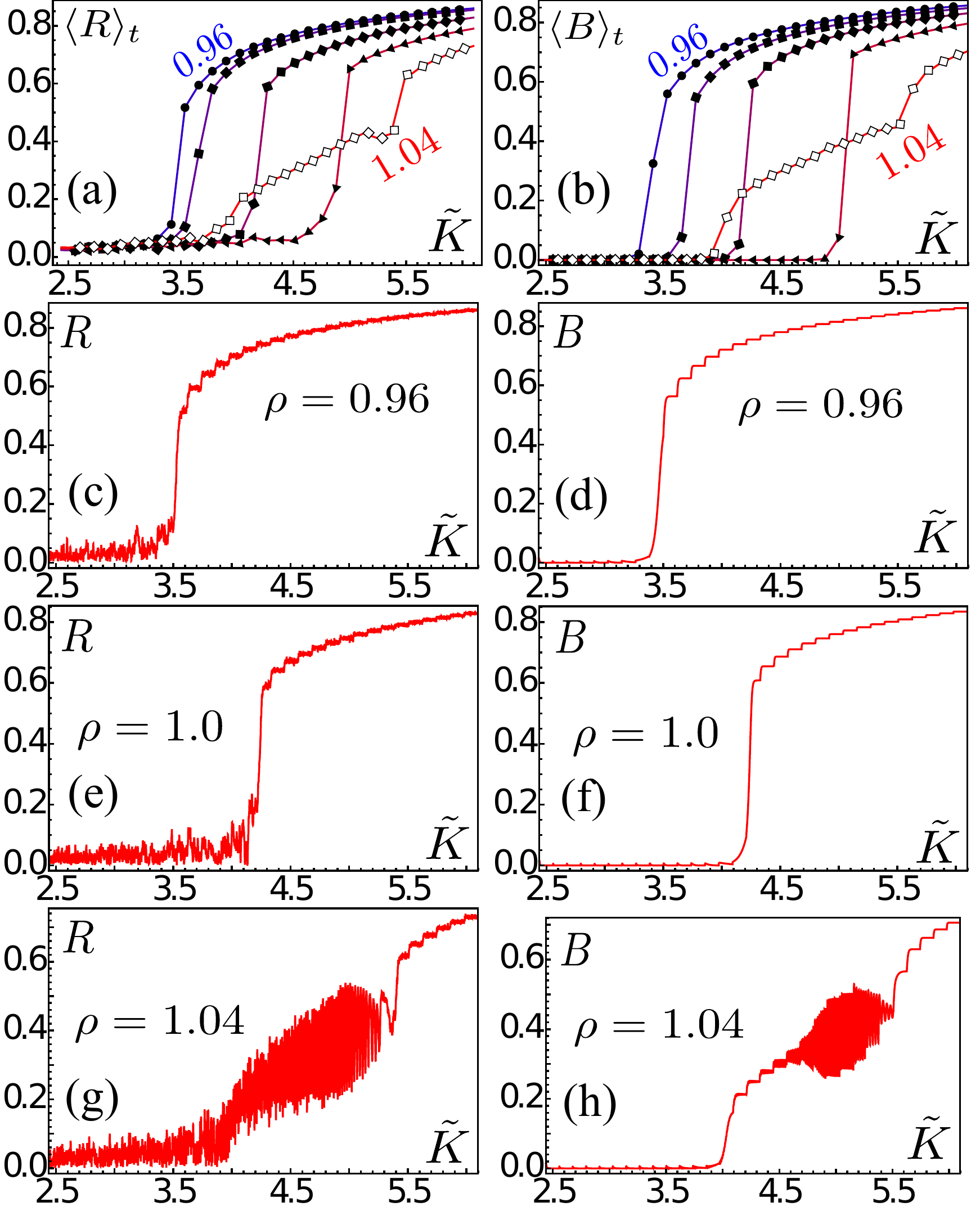}
    \caption{Top left: Time-averaged value of the order parameter $R(t)$ calculated directly from Eqs.~(1). Top right: Time-averaged value of the order parameter $B(t)$ calculated using the reduced equations (13). For both top panels, results are shown for $\rho = 0.96$ (circles), $\rho = 0.98$ (squares), $\rho = 1.00$ (diamonds), $\rho = 1.02$ (triangles), and $\rho = 1.04$ (inverted triangles). The lower three rows show the timeseries $R(t)$ (left) and $B(t)$ (right) for $\rho = 0.96$ (top), $\rho = 1.00$ (middle), and $\rho = 1.04$ (bottom). Note that we defined $\tilde K \equiv 100 K$.}\label{panel}
\end{figure}

In Fig.~\ref{panel} we compare the results of directly solving (1) (left column) with those obtained from solving the reduced equations (13) (right column). In Fig.~\ref{panel}~(a) we show the time-averaged value of the order parameter $R(t)$ for $\rho = 0.96$ (circles), $\rho = 0.98$ (squares), $\rho = 1.00$ (diamonds), $\rho = 1.02$ (triangles), and $\rho = 1.04$ (small empty diamonds), and in Fig.~\ref{panel} (b) we show the time-averaged value of the order parameter $B(t)$  for the same values of $\rho$. In the lower three rows we show the timeseries of $R(t)$ (left) and $B(t)$ (right) for $\rho = 0.96$ (top), $\rho = 1.0$ (middle), and $\rho =1.04$ (bottom).

The reduced theory reproduces the behavior observed in the simulations of the full model. In particular, both methods show a nontrivial dependence of the critical coupling constant $K_c$ on the assortative coefficient $\rho$. For networks in which the frequency is not correlated with the node's degrees, $K_c$ is inversely proportional to $\rho$ when $\rho$ is close to $1$ [see Eqs.~(17) and (18)]. In our example, however, $K_c$ increases with $\rho$ for $\rho < 1.02$ and decreases for $\rho > 1.02$. Our reduced model can be used to understand this dependence. In the next Section, we use Eqs.~(13) to derive an implicit equation for $K_c$ valid for small assortativity ($\rho \approx 1$), which predicts the nonmonotonic behavior observed in the simulations.

Referring to the three lower rows of Fig.~\ref{panel}, we see that, as claimed for the example in the paper, the lower values of $\rho$ [corresponding to row 2 ($\rho = 0.96$, disassortative) and row 3 ($\rho = 1.0$)] show a bifurcation from incoherence ($R$ and $B \approx 0$) to a steady behavior that persists as $K$ is increased, while at sufficiently large positive assortativity [corresponding to row 4 ($\rho = 1.04$)] bifurcations involving oscillatory states become possible. This is most clearly seen in the right panel of the bottom row which shows the behavior of $B(t)$ obtained from the numerical solution of our reduced description. After bifurcating from incoherence to a steady state at $\tilde K \approx 4.0$ (recall that for this example we defined $\tilde K \equiv 100K$), there appears to be a Hopf bifurcation to periodic oscillations of $B(t)$ occurring at about $\tilde K \approx 4.6$, followed by a return to steady motion at about $\tilde K \approx 5.5$. Blowing up the behavior around $\tilde K \approx 5.5$ (not shown), as in Figs.~5(d) and 5(e), we see that the period apparently diverges at $\tilde K \approx 5.5$, as occurs for a SNIPER bifurcation. Referring now to the left panel of the bottom row, which shows $R(t)$ obatined from numerical solution of Eq.~(1), we see that, although (as discussed in the main text) there is substantial noise (particularly near $\tilde K \approx 4.0)$ and also that there are quantitative differences with the $B$ versus $K$ plot (on the right), it is nevertheless striking that the qualitative fewatures of an onset of oscillations and of the apparent divergence of the oscillation period (at about $\tilde K \approx 5.5$), characteristic of a SNIPER bifurcation, are observed in both the $R(t)$ and $B(t)$ bottom row ($\rho = 1.04$) plots.

\section{Perturbative expression for $K_c$ for ${\bf k}$ dependent $\omega_0$ and $\Delta$}\label{kkc}

In this Section we derive an equation to determine $K_c$ in the presence of small assortativity and ${\bf k}$ dependence of our Lorenzian $g(\omega|{\bf k})$, i.e., $\omega_0({\bf k})$ and $\Delta({\bf k})$ are ${\bf k}$ dependent. Assuming a steady synchronized solution of the form $\hat b({\bf k},t) = \bar b({\bf k}) e^{-i \Omega t} $ in Eq.~(13), we obtain
\begin{equation}\tag{S9}
[-i\{\omega_0({\bf k}) +\Omega\} + \Delta({\bf k})]\bar b({\bf k}) = \frac{K}{2} \sum_{{\bf k}'} P({\bf k}')a({\bf k}' \to {\bf k}) \bar b({\bf k}').
\end{equation}
Letting $D({\bf k}) \equiv [-i\{\omega_0({\bf k}) +\Omega\} + \Delta({\bf k})]$ and $c({\bf k}) \equiv D({\bf k})\bar b({\bf k})$, we get
\begin{equation}\label{b2}\tag{S10}
c({\bf k}) = \frac{K}{2} \sum_{{\bf k}'} \frac{P({\bf k}')}{D({\bf k}')}a({\bf k}' \to {\bf k})c({\bf k}').
\end{equation}
When there is no assortativity, 
$$a({\bf k}' \to {\bf k}) = \frac{{k^{out}}' k^{in}}{N\langle k \rangle},$$
so the lowest order result is
\begin{equation}\tag{S11}
c^{(0)}({\bf k}) = \frac{K}{2} \sum_{{\bf k}'} \frac{P({\bf k}')}{D({\bf k}')} \frac{{k^{out}}' k^{in}}{N\langle k \rangle}c^{(0)}({\bf k}'),
\end{equation}
i.e., the zero order right eigenvector is $c^{(0)}({\bf k}) = (\mbox{const.}) k^{in}$ and the zero order left eigenvector is $k^{out} P({\bf k})/D({\bf k})$. Inserting the right eigenvector $c^{(0)}$ in (S10), multiplying by the zero order left eigenvector, and summing over ${\bf k}$, we get the first order result,
\begin{equation}
\sum_{{\bf k}} \frac{k^{in} k^{out} P({\bf k})}{D({\bf k})} = \nonumber
\end{equation}
\begin{equation}\label{eqkc}\tag{S12}
\frac{K_c}{2} \sum_{{\bf k}} \sum_{{\bf k}'} \frac{k^{out} P({\bf k})}{D({\bf k})} a({\bf k}' \to {\bf k}) \frac{{k^{in}}' P({\bf k}')}{D({\bf k}')}
\end{equation}

\begin{figure}[t]
  \centering
   \includegraphics[width=0.9\linewidth]{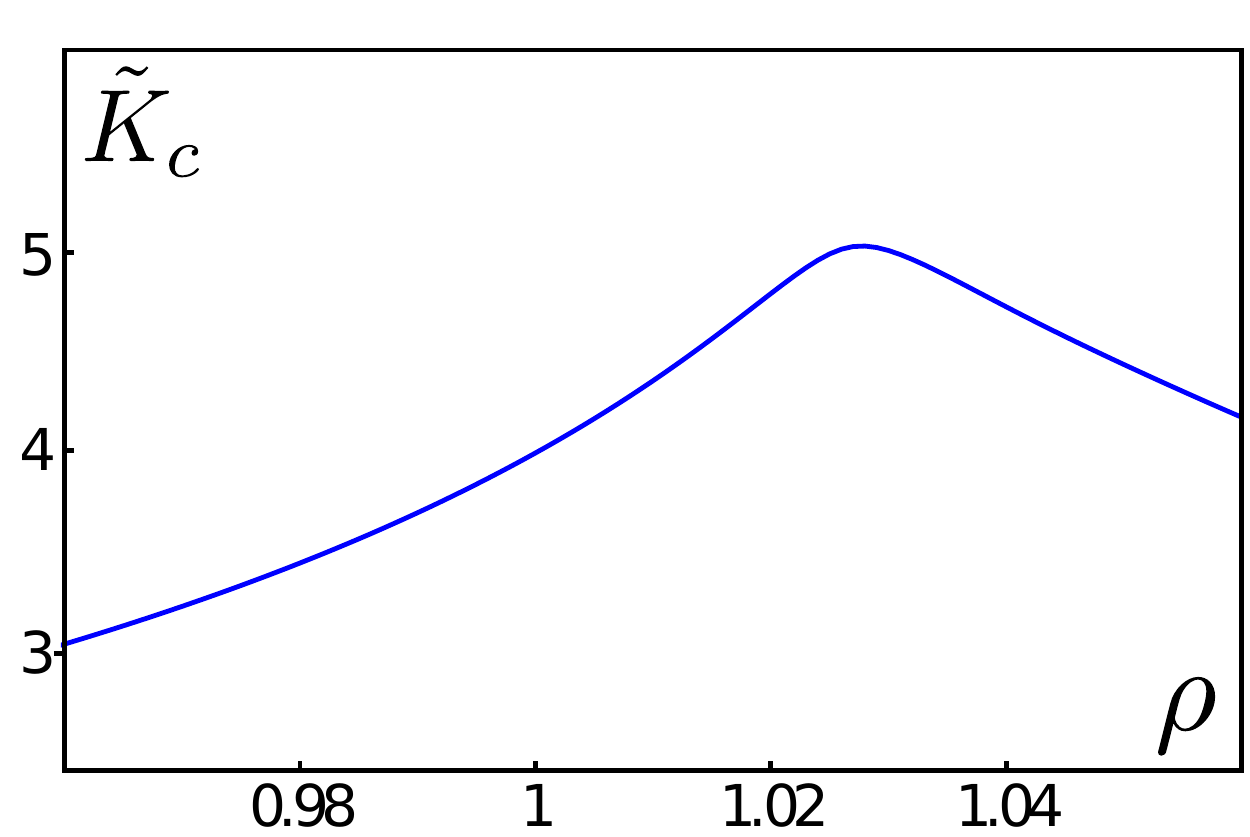}
    \caption{Critical coupling constant $\tilde K_c = 100 K_c$ as a function of $\rho$ predicted from numerical solution of Eq.~(\ref{eqkc}).}\label{teokc}
\end{figure}

which, together with $D({\bf k}) \equiv [-i\{\omega_0({\bf k}) +\Omega\} + \Delta({\bf k})]$, gives one complex equation for the two real unknowns $K_c$ and $\Omega$.
Solving this equation numerically gives the plot shown in Figure~\ref{teokc}. The predicted value of $K_c$ agrees with the observed value at $\rho = 1$ ($100K_c \approx 4.0$) and agrees qualitatively elsewhere. In particular, the plot shows the same nonmonotonic dependence of $\tilde K_c$ on $\rho$ observed in the simulations (e.g., top panels of Fig.~\ref{panel}).


\end{document}